\pdfoutput=1
\documentclass[aps,prl,10pt,twocolumn,showpacs,superscriptaddress]{revtex4-1}
\usepackage{amsmath}
\usepackage{latexsym}
\usepackage{pgfplots}
\usepackage{rotating}
\usepackage{amssymb}
\usepackage{bm}
\usepackage{graphics,epstopdf}
\usepackage{color}\usepackage{amsmath}

\usepackage[usenames,dvipsnames,svgnames]{pstricks}
\usepackage{epsfig}
\usepackage{pst-grad} 
\usepackage{pst-plot} 

\usepackage{newlfont}
\usepackage{amssymb}
\usepackage{amsfonts}
\usepackage{amsmath}
\usepackage{mathrsfs}
\usepackage{amsthm}
\usepackage{graphicx}
\usepackage{epsfig}
\usepackage{color}
\usepackage{bm}
\usepackage[breaklinks]{hyperref}
\usepackage{tikz}
\newcommand{\ket}[1]{|{#1}\rangle}
\newcommand{\bra}[1]{\langle{#1}|}

\usepackage{times}

\begin{document}

\title{Distinguishing two preparations for same pure state leads to signalling}

\author{Arun Kumar Pati }
\affiliation{Quantum Information and Computation Group,\\
Harish-Chandra Research Institute, Chhatnag Road, Jhunsi, Allahabad, India}


\date{\today}

\begin{abstract}
Pure state of a physical system can be prepared in an infinite number of ways. Here, we prove that given a pure state of a quantum system it is impossible to 
distinguish two preparation procedures. Further, we show that if we can distinguish two preparation procedures for the same pure state then that can lead to signalling.
This impossibility result is different than the no measurement without disturbance and the no-cloning. Extending this result for a  pure bipartite entangled 
state entails that the impossibility of distinguishing two preparation procedures for a mixed state follows from the impossibility of distinguishing
two preparations for a pure bipartite state.
\end{abstract}

\maketitle

  In quantum mechanics, the state of a physical system, be it pure or mixed, is supposed to capture the complete description of a system. 
  It is known that if we describe the system
  by a mixed state, then the same mixed state can be prepared in an infinite number of ways by probabilistic mixing of different decompositions of the same 
  density matrix. Once the state is prepared, it is impossible in principle to
  distinguish two (or more) preparation procedures for a mixed state. In fact, if we can distinguish two preparations for a mixed state, then we could have signalling and we could 
  violate the Second law of thermodynamics \cite{peres}.
  
  If a physical system is in a pure state, that can also be prepared in an infinite number of ways. Since a Hilbert space can have an infinite number of orthonormal 
  basis sets, we can expand a pure state using any orthonormal basis. Consider two observables $A$ and $B$ with eigenbasis sets $\{\ket{a_n} \}$ and $\{\ket{b_n} \}$, 
  respectively. Then, we know that we can express the pure state using the eigenbasis of the observables $A$, i.e., $\ket{\psi} = \sum_n \alpha_n \ket{a_n}$ or we could also 
  expand the pure state using the eigenbasis of the observables $B$, i.e., $ \ket{\psi} = \sum_n \beta_n \ket{ b_n} $. Each of these possible expansions represents one 
  possible preparation procedure. In quantum mechanics, we can prepare a pure state starting from a fiducial pure state either by a unitary transformation,  or by a 
  general quantum operation. Since this has to be a physical operation, and any physical operation can be realized by a unitary transformation of the fiducial state along with an 
  ancillary state, without loss of generality, we consider here only unitary operation as a physically realizable process to prepare a pure state.
  
  To convey the main result, we consider a quantum system in a two-dimensional Hilbert space and one can generalize our results to higher dimensional systems.
  For example, consider a qubit in a state $\ket{\psi} = \alpha \ket{0} + \beta \ket{1}$. This can be prepared starting from an initial state 
  $\ket{0}$ (say along up-$z$-axis of a spin-half particle) 
  and by applying a unitary $U(\alpha, \beta)$, i.e., $\ket{\psi} = U(\alpha, \beta) \ket{0}$. The same state can also be prepared starting from an initial state
  $\ket{+}$ (say up-$x$-axis of a spin-half particle) and by applying a different unitary $V(\alpha, \beta)$, i.e., $\ket{\psi} = V(\alpha, \beta) \ket{+}$. 
  The question we address is whether it is possible to distinguish these two preparation procedures by any physical operation. To our surprise, we
  discovered that the impossibility of distinguishing two preparation procedures for a pure state has never been proved in the literature, not to mention 
  about how that can lead to signaling. Here, we show that it is impossible to distinguish two (or more) preparations for a pure state. Then, we prove that 
  if we can distinguish two different preparations for a pure state, then one 
  can have signaling. Also, we argue that the new impossibility result is different than the other no-go theorems such as the impossibility of 
  measuring without disturbance \cite{park} and the no-cloning in quantum information \cite{wz,dd}.  Towards the end, we prove that the impossibility of distinguishing two 
  preparation procedures for a mixed state follows from the impossibility of distinguishing
two preparations for a pure bipartite state. 

 Suppose that there is a machine which can distinguish two preparation procedures for a pure state. If the machine respects quantum mechanics, then that has to be 
 represented by a physical operation. We can always realize a physical operation as a unitary evolution on a larger Hilbert space. Imagine that the physical operation
 is a unitary operation on the system and the machine state. Now, assume that the machine somehow knows the preparation procedure and the final state of the machine changes
 according to the preparation procedure. Thus, the unitary transformation that may distinguish two preparations for a qubit state is given by

 \begin{eqnarray}
 U\ket{0} \ket{A} & \rightarrow  \ket{\psi} \ket{A_{U_0} }, \nonumber \\
 V\ket{+} \ket{A} & \rightarrow  \ket{\psi} \ket{A_{V_+} }, 
 \end{eqnarray}
where $\ket{A }$ is the initial state of the machine, $\ket{A_{U_0} }$ is the final state of the machine if the pure state is prepared via $U\ket{0}$ and 
$\ket{A_{V_+} }$ is the final state of the machine if the pure state is prepared via $V\ket{+}$. The `preparation-distinguishing' machine, if it exists, then 
that will change the final state of the machine according to the preparation procedure and leaves the input state unchanged. Note that we need the states 
$\ket{A_{U_0} }$ and $\ket{A_{V_+} }$ to be different, so that we can obtain information about the preparation procedures of the pure state.
However, if this is to hold, then we have 

\[
1 = \bra{A_{U_0} } \ket{A_{V_+} }.
\]
This implies that $\ket{A_{U_0} }$ and  $\ket{A_{V_+} }$ can never be different and hence there is no way to distinguish two preparation procedures for a 
pure state.

 Note that the machine that we have defined above is distinct one compared to the machine that is supposed to measure two non-orthogonal quantum states without disturbance 
 In this case, the transformation is defined as
 
 \begin{eqnarray}
 \ket{0} \ket{A} & \rightarrow  \ket{0} \ket{A_0 } \nonumber \\
\ket{+} \ket{A} & \rightarrow  \ket{+} \ket{A_+}.
 \end{eqnarray}
That this process is also impossible follows from the unitarity, because we can never be able to satisfy $1 = \bra{A_0 } \ket{A_+ }$. 
This is paraphrased by saying that `it is impossible to distinguish two non-orthogonal states without disturbance'.
However, note that these two machines are completely different.
This is because there is no way that we can go from Eq(1) to Eq(2) as the unitaries $U$ and $V$ in general will not commute with the global operation that realizes 
the process given in Eq(1). 
Therefore, the hypothetical machine that can distinguishing two preparation procedures for a pure state is fundamentally different than the machine that is supposed to 
distinguish two non-orthogonal quantum  states without disturbance. Therefore, this impossibility is a new result independent of the earlier one. We can also argue that it is
independent of the no-cloning theorem \cite{wz,dd}. First, note that if we know the complete preparation procedure, then we know the state of a qubit. But the converse is not true, i.e., 
knowing the state of a qubit is not same as knowing the preparation procedure. If we know the state of a qubit, then we can clone it, whereas here, even if we know the state
we cannot distinguishing two preparation procedures. This shows that the present no-go theorem is different from the other no-go theorems such as the 
no-cloning  \cite{wz,dd} and the no-deleting theorems \cite{pati}.

{\it Distinguishing two preparations for pure state and signalling:}
Here, we will show that distinguishing two different preparations for a pure state can actually lead to signalling.
Imagine that Alice and Bob share an entangled Einstein-Podolsky-Rosen (EPR) pair as given by 
 \begin{eqnarray}
 \ket{\Psi^{-}} = \frac{1}{\sqrt 2} ( \ket{0}\ket{1} - \ket{1}\ket{0}).
\end{eqnarray}
The EPR state satisfies the property 
\begin{eqnarray}
 \ket{\Psi^{-}} = U(\alpha, \beta) \otimes U(\alpha, \beta) \ket{\Psi^{-}} = \frac{1}{\sqrt 2} ( \ket{\psi}\ket{{\bar \psi} } - \ket{{\bar \psi} }\ket{\psi}), \nonumber \\
\end{eqnarray}
where $\ket{\psi} = \alpha \ket{0} + \beta \ket{1} = U(\alpha, \beta) \ket{0} $ and $\ket{{\bar \psi} } = \alpha^* \ket{1} - \beta^* \ket{0} = U(\alpha, \beta) \ket{1} $.
This invariance property of singlet is equivalent to
\begin{eqnarray}
 U^{\dagger}(\alpha, \beta) \otimes I \ket{\Psi^{-}} & =  I  \otimes U(\alpha, \beta) \ket{\Psi^{-}} \nonumber \\
 & = \frac{1}{\sqrt 2} ( \ket{0} U \ket{1} - \ket{1}U \ket{0}). 
 \end{eqnarray}
Physically, this means that if Alice applies $U^{\dagger}(\alpha, \beta)$ on her particle this is equivalent to applying $U(\alpha, \beta)$ on Bob's particle.
Similarly, the invariance property for the singlet implies that we have 
\begin{eqnarray}
 V^{\dagger}(\alpha, \beta) \otimes I \ket{\Psi^{-}} & =  I  \otimes V(\alpha, \beta) \ket{\Psi^{-}} \nonumber \\ 
 & = \frac{1}{\sqrt 2} ( \ket{+} V \ket{- } - \ket{-} V \ket{+}), 
\end{eqnarray}
with the notion that $\ket{\psi} = V(\alpha, \beta) \ket{+}$ and $\ket{{\bar \psi}} = V(\alpha, \beta) \ket{-}$.
Now, let us encode one classical bit in Alice's action, i.e., if she receives $0$, then she applies $U^{\dagger}(\alpha, \beta)$ on her particle and if she receives 
$1$, then she applies $V^{\dagger}(\alpha, \beta)$ on her particle. These two choices by Alice allow us to have the possibility of two different preparations at Bob's end. 
Now assume that Bob has a hypothetical 
machine which can distinguish two preparation procedures for a pure state. Bob attaches the machine and allows the transformation as given by 

\begin{eqnarray}
 U\ket{0} \ket{A} & \rightarrow  \ket{\psi} \ket{A_{U_0} }, ~~~~ V\ket{+} \ket{A} & \rightarrow  \ket{\psi} \ket{A_{V_+} }, \nonumber \\
 U\ket{1} \ket{A} & \rightarrow  \ket{{\bar \psi}} \ket{A_{U_1} }, ~~~~ V\ket{-} \ket{A} & \rightarrow  \ket{{\bar \psi} } \ket{A_{V_-} }.
 \end{eqnarray}
 Then, depending on the two 
choices of preparations of a pure state, we have 
\begin{align}
  \frac{1}{\sqrt 2} ( \ket{0} U \ket{1 } \ket{A}  - \ket{1}U \ket{0} \ket{A})  &\rightarrow \nonumber \\
 & \frac{1}{\sqrt 2} ( \ket{0}\ket{{\bar \psi} } \ket{A_{U_1} }  - \ket{1}\ket{\psi} \ket{A_{U_0}}), \nonumber \\
 \frac{1}{\sqrt 2} ( \ket{+} V \ket{- } \ket{A} - \ket{-} V \ket{+} \ket{A} )   & \rightarrow  \\
 &\frac{1}{\sqrt 2} ( \ket{0}\ket{{\bar \psi} } \ket{A_{V_-} }  - \ket{1}\ket{\psi} \ket{A_{V_+}}). \nonumber \\
\end{align}
Now, the two preparation procedures gives two different density matrices at Bob's end. These are given by 

\begin{eqnarray}
\rho_B^{0} =  \frac{1}{2} [ \ket{{\bar \psi} }  \bra{{\bar \psi} } \otimes \ket{A_{U_1} }   \bra{A_{U_1} }   +  \ket{\psi}  \bra{\psi} \otimes \ket{A_{U_0}} \bra{A_{U_0}} ], \nonumber \\
\rho_B^{+} =  \frac{1}{2} [ \ket{{\bar \psi} }  \bra{{\bar \psi} } \otimes \ket{A_{V_-} }   \bra{A_{V_-} }   +  \ket{\psi}  \bra{\psi} \otimes \ket{A_{V_+}} \bra{A_{V_+}} ], \nonumber \\
\end{eqnarray}
where $\rho_B^{0}$ is the result of one preparation procedure and $\rho_B^{+}$ is the result of other preparation procedure.
Since these two density matrices are different, Bob can infer Alice's action, thus revealing one bit of information without any communication from
Alice. This would lead to signaling. Therefore, from the no-signaling, we can argue that it is impossible to distinguish two preparation procedures for a pure state.

{\it Impossibility of distinguishing two preparations for bipartite states}: 
Now, we ask can the impossibility of distinguishing two preparations for a pure bipartite state lead to the impossibility of distinguishing preparations for a mixed state?
In quantum mechanics, a density matrix can have infinite number of decompositions (proper mixtures) and it is impossible to distinguish two preparation procedures.
Also, we know that a mixture (improper) occurs when we trace out one of the subsystem of an entangled state. 
In what follows, we show that the impossibility to distinguish two preparations for a pure bipartite entangled state indeed 
implies the impossibility of distinguishing two preparation procedures for a mixed state.

First, note that a pure bipartite entangled state can also be prepared in an infinite number of ways. Consider two  preparation procedures of a pure bipartite state 
$\ket{\Psi}_{AB}$ which are expressed as $\sum_{nm} C_{nm} \ket{\psi_n} \ket{\phi_m}$ and $ \sum_{\mu \nu} \alpha_{\mu \nu} \ket{a_{\mu} } \ket{b_{\nu}}$. 
Extending our earlier result to a pure 
 bipartite state, we can show that it is impossible to distinguish two different preparations. Now, these two 
 preparation procedures for a pure state will result in two possible preparations for the density matrix of either subsystem. For example, if we trace out 
 the second subsystem, we will have density matrix $\rho_A = \sum_m \ket{{\tilde \psi}_m} \bra{{\tilde \psi}_m}$, 
 where $\ket{{\tilde \psi}_m} = \sum_n C_{nm} \ket{ \psi_n}$ are  unnormalized and non-orthogonal states with $\sum_m || {\tilde \psi}_m ||^2 = 1 $. 
Similarly, for the other preparation procedure, if we trace out the second subsystem, then the density matrix $\rho_A = \sum_{\nu}  \ket{{\tilde a}_{\nu}} \bra{{\tilde a}_{\nu}}$, 
 where $\ket{{\tilde a}_{\nu} } = \sum_{\mu} \alpha_{\mu \nu} \ket{ a_{\mu} }$ are unnormalized and non-orthogonal states with $\sum_{\nu} || {\tilde a}_{\nu} ||^2 =1  $. 

  Now, suppose that there is a physical operation that can perfectly distinguish two preparations for the same mixed state. This means that 
  the `preparation-distinguishing' machine can result 
  in two different states of the system and the machine as $\rho_{AE}^{(1)}$ and  $\rho_{AE}^{(2)}$ corresponding to two different preparation procedures, respectively,
  with $D(\rho_{AE}^{(1)}, \rho_{AE}^{(2)} ) =1$.  Here $D$ is a measure of distinguishing two different 
  preparations which are labeled as $`1'$ and $`2'$. This physical operation can also be realized on a purified 
  Hilbert space that results in two possible states $\ket{\Psi_{ABE}^{(1)} }$  and $\ket{\Psi_{ABE}^{(2)} }$  according to two different preparation procedures. 
  Note that the distinguishabilty measure decreases under partial tracing, therefore, we have $D(\rho_{AE}^{(1)}, \rho_{AE}^{(2)} ) \le D(\Psi_{ABE}^{(1)}, \Psi_{ABE}^{(2)} )  $. 
  Since we have assumed that $D(\rho_{AE}^{(1)}, \rho_{AE}^{(2)} ) =1$, this then implies 
  that  $D(\Psi_{ABE}^{(1)}, \Psi_{ABE}^{(2)} ) \ge 1$. If the distinguishabilty measure satisfies $ 0 \le D \le 1$, then it must be true that 
  $D(\Psi_{ABE}^{(1)}, \Psi_{ABE}^{(2)} ) = 1$, i.e., we can distinguish two preparations for a pure bipartite entangled state perfectly. 
  But we know that we cannot distinguish two different preparations for the same pure entangled state and hence it is impossible to
distinguish two preparations for the same mixed state.

{\it Conclusions:}
In quantum theory, the preparation of a physical system and the measurement procedure play fundamental role. Measurement process though always entails an outcome 
which may be random, an experimentalist must be able to reproduce the preparation and measurement procedures. A reproducible preparation of a physical system is 
represented by a pure state (in the case of closed system) or by a density operator (in the case of open system).  There are infinite number of ways in which a given pure 
state can be prepared and hence there is an 
infinite number of pasts associated to a present pure state of a physical system. Our results shows that once the state is prepared in a 
pure state, then there is no way to reveal its preparation procedure.
We have also shown that the violation of the impossibility of distinguishing
two different preparations for a pure state can lead to signalling. Moreover, we have argued that our result is independent of the no measurement without disturbance 
and the no-cloning. We have also proved that the impossibility of distinguishing two preparations for the same mixed state follows from a more fundamental result
that it is impossible to distinguish two different preparations for the same pure bipartite state.
The new impossibility result, hitherto unnoticed, has a different status compared to other no-go theorems such as the no-cloning and the no-deleting theorems 
in quantum information, and opens up several questions of fundamental importance in quantum theory.

\vskip 1cm

\noindent
{\it Note}: I thank Ujjwal Sen and Chirag Srivastva for useful discussions. For another perspective on distinguishing and signalling 
readers can see today's paper in the arXiv by Srivastva {\it et al}.

\bibliographystyle{h-physrev4}

\end{document}